\documentclass[11pt]{article}
\usepackage{moriond,epsfig}

\def\Journal#1#2#3#4{{#1} {\bf #2}, #3 (#4)}

\def\PLB{{\em Phys. Lett.}  B}
\def\PRL{\em Phys. Rev. Lett.}

\def\be{\begin{equation}}
\def\ee{\end{equation}}
\def\bea{\begin{eqnarray}}
\def\eea{\end{eqnarray}}

\begin{document}
\vspace*{4cm}
\title{SPECTROSCOPIC MEASUREMENTS USING THE H1 AND ZEUS DETECTORS}

\author{ S. SCHMIDT }

\address{Deutsches Elektronen-Synchrotron, Notkestra\ss e 85,\\
22607 Hamburg, Germany}

\maketitle\abstracts{
Results on spectroscopy from the H1 and ZEUS collaborations are presented. The main focus is to search for
baryon states which could be interpreted as pentaquarks. This includes states decaying
to $K_s^0 p$ and $K_s^0\bar{p}$, $\Xi\pi$ and $D^\ast p$. In addition an analysis of $K^0_sK^0_s$ 
resonances is covered.}

\section{Introduction}

At the $ep$-collider HERA, interactions are studied at a centre-of-mass energy of 
300--320\,GeV by the H1 and ZEUS multi-purpose detectors. The virtuality, 
$Q^2$, of the exchanged photon allows to distinguish two kinematical regimes: photoproduction ($Q^2 < 1\,$GeV$^2$) and 
deep-inelastic-scattering (DIS) ($Q^2 > 1\,$GeV$^2$). As recent results from fixed-target 
experiments have evidence for a narrow
baryon resonance decaying to $K^+ n$ \cite{pq1} and $K_s^0p$ \cite{pq2}, interpreted as a 
pentaquark, spectroscopic measurements covering this and related topics have been performed at HERA.

\section{Evidence for a narrow baryonic state decaying to $K_s^0 p$ and $K_s^0\bar{p}$
in DIS}

A resonance search\cite{z1} has been made in the $K_s^0p$ and $K_s^0\bar{p}$
invariant-mass spectrum measured with the ZEUS detector using
an integrated luminosity of 121 pb$^{-1}$. The search was performed in
the central rapidity region of inclusive DIS for $Q^2$ above 1\,GeV$^2$.
The results support the existence of a state like those observed by the fixed-target 
experiments, with a mass of $1521.5\pm 1.5$ (stat.) $^{+2.8}_{-1.7}$ (syst.)\,MeV 
and a Gaussian width consistent with the experimental resolution of
$2\,$MeV. The signal is visible at moderate $Q^2$ and, for $Q^2 > 20\,$GeV$^2$, contains $221 \pm 48$
events (Fig. 1 (a)). The probability of a similar signal anywhere 
in the range 1500--1560\,MeV arising from fluctuations of the background is below 
$6 \times 10^{-5}$.

\begin{figure}
\begin{picture}(200,230)
\put(0,2){\includegraphics[scale=0.27]{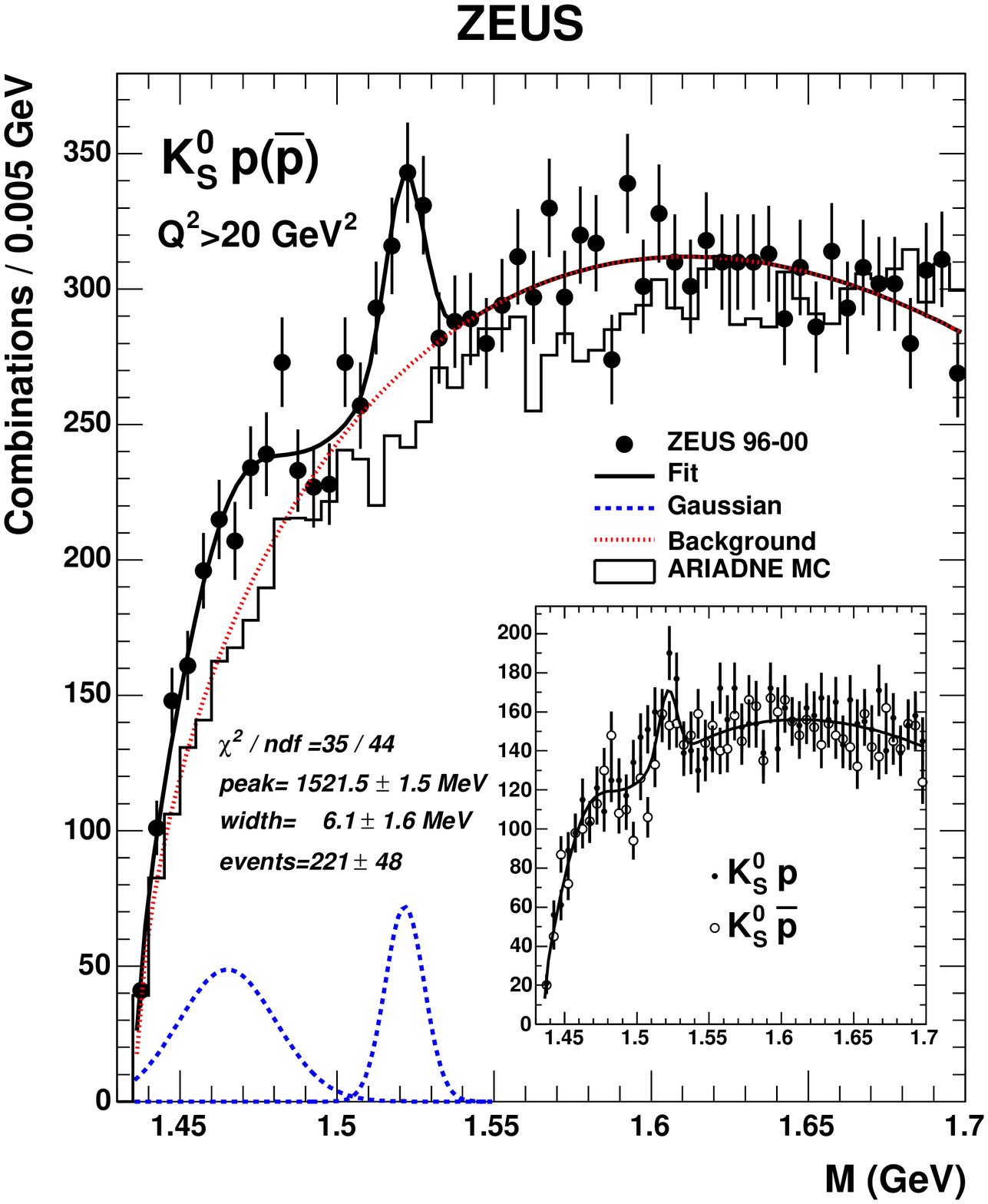}}
\put(170,0){\includegraphics[scale=0.50]{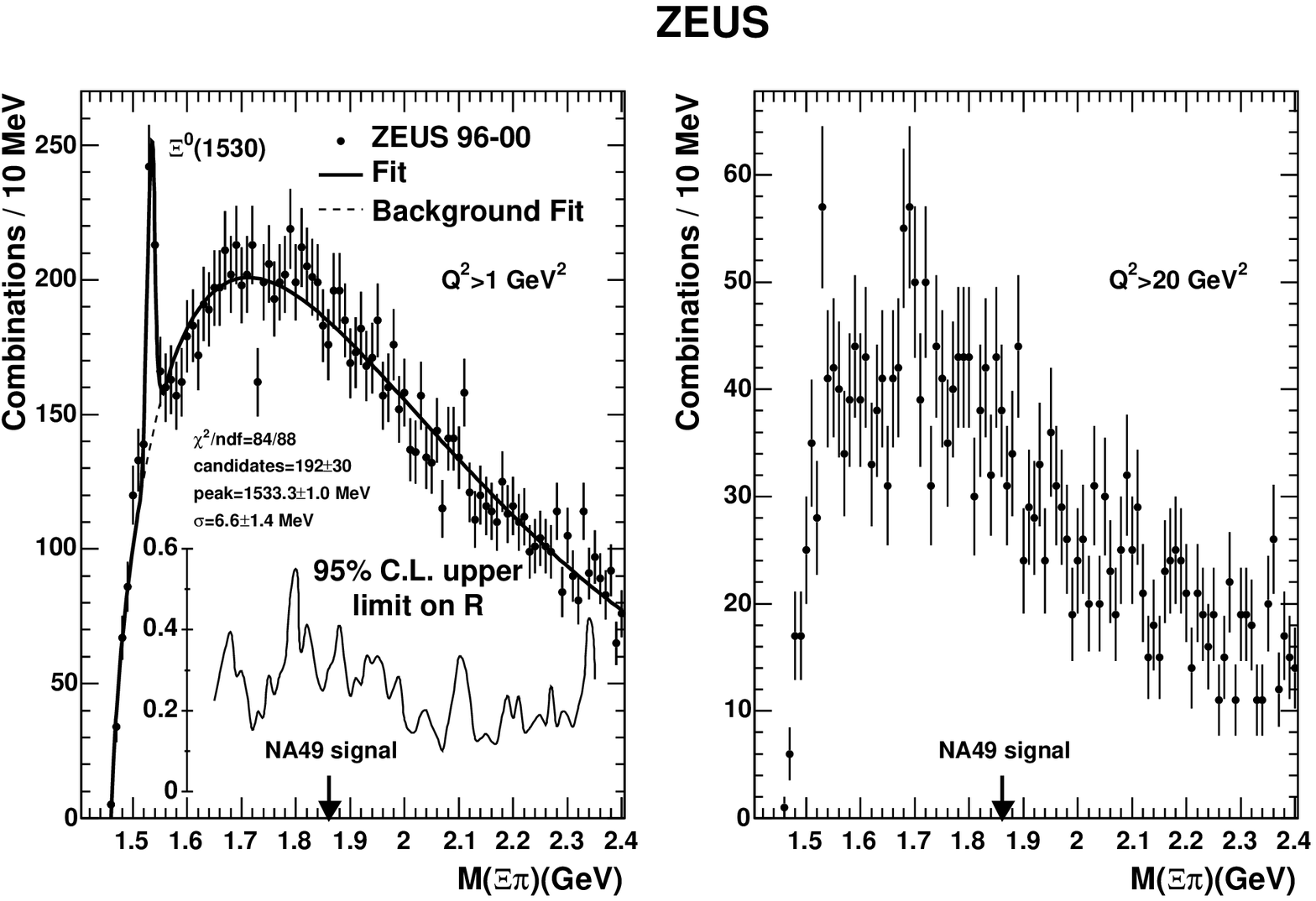}}
\put(130,158){\tiny(a)}
\put(285,158){\tiny(b)}
\put(429,158){\tiny(c)}
\end{picture}
\label{fig:DESY-04-056_4}
\caption{(a) Invariant-mass spectrum for the $K_s^0p(\bar{p})$ channel for $Q^2 > 20\,$GeV$^2$. The solid
line is the result of a fit using two Gaussians plus background parametrisation. The histogram
shows the prediction of the {\sc ARIADNE} Monte Carlo, the inset the charge separated samples. (b) The
$\Xi\pi$ invariant-mass spectrum for $Q^{2}>1\,$GeV$^2$ and (c) for $Q^{2}>20\,$GeV$^{2}$. All four charge
combinations have been summed up. The solid line in (b) is the result of a fit to the data using a Gaussian
and a parametrised background. Also shown is the  $95\%$ C.L. upper limit on $R$, the ratio of the 
$\Xi^{--}_{3/2}$($\Xi^{0}_{3/2}$) signal to $\Xi^{0}(1530)$ as a function of the invariant mass for
$Q^{2}>1\,$GeV$^{2}$. The arrows show the location of the signal observed by NA49.}
\end{figure}

\section{Search for pentaquarks decaying to $\Xi\pi$ in DIS}

The ZEUS collaboration has also performed a search\cite{z2} for pentaquarks decaying
to $\Xi^{-}\pi^{-}$ ($\Xi^{-}\pi^{+}$) and corresponding antiparticles using 
the same data sample as in the previous analysis, to investigate the observation of the
NA49 experiment\cite{na49}.
A clear signal for $\Xi^{0}(1530)\rightarrow\Xi^{-}\pi^{+}$
was observed. However, no signal for any new baryonic state was observed at 
higher masses in either the $\Xi^{-}\pi^{-}$ or $\Xi^{-}\pi^{+}$ channels (Fig. 1
(b) and (c)). The searches in the antiparticle channels were also negative. Upper limits on the
ratio of a possible $\Xi^{--}_{3/2}$ ($\Xi^{0}_{3/2}$) signal to the $\Xi^{0}(1530)$
signal were set in the mass range 1650--2350 MeV.

\section{Evidence for a narrow anti-charmed baryon state by H1}

The H1 collaboration observed\cite{h1} a narrow resonance in $D^{\ast-}p$ and $D^{\ast+}\bar{p}$ 
invariant-mass combinations in a data sample corresponding to an integrated luminosity of $75\,$pb$^{-1}$ in DIS.
The decay channel $D^{\ast+}\rightarrow D^0\pi^{+}_{s}\rightarrow(K^{-}\pi^{+})\pi^{+}_{s}$ (and the 
corresponding antiparticle decay) was used to identify the $D^{\ast+}$ mesons, the (anti-)proton candidates
were selected using particle identification based 
on the differential energy loss. The resonance has a mass of $M(D^\ast p)=3099\pm3$\,(stat.)\,$\pm5$ 
(syst.)\,MeV and a measured Gaussian 
width of $12\pm3$\,(stat.)\,MeV, compatible with the experimental resolution (Fig. 2 (a)). The probability for a
background fluctuation to produce a signal as large as observed is less than $4 \times 10^{-8}$. The
region of $M(D^\ast p)$ in which the signal is observed contains a richer yield of $D^\ast$ mesons
and exhibits a harder proton candidate momentum distribution than is the case for the side bands in 
$M(D^\ast p)$. The fraction of $D^\ast$ mesons originating from the resonance decay is found to be $1.46\pm0.32\%$.
A signal with compatible mass and width is also observed in an independent photoproduction sample.

\begin{figure}
\begin{picture}(200,230)
\put(0,0){\includegraphics[scale=0.40]{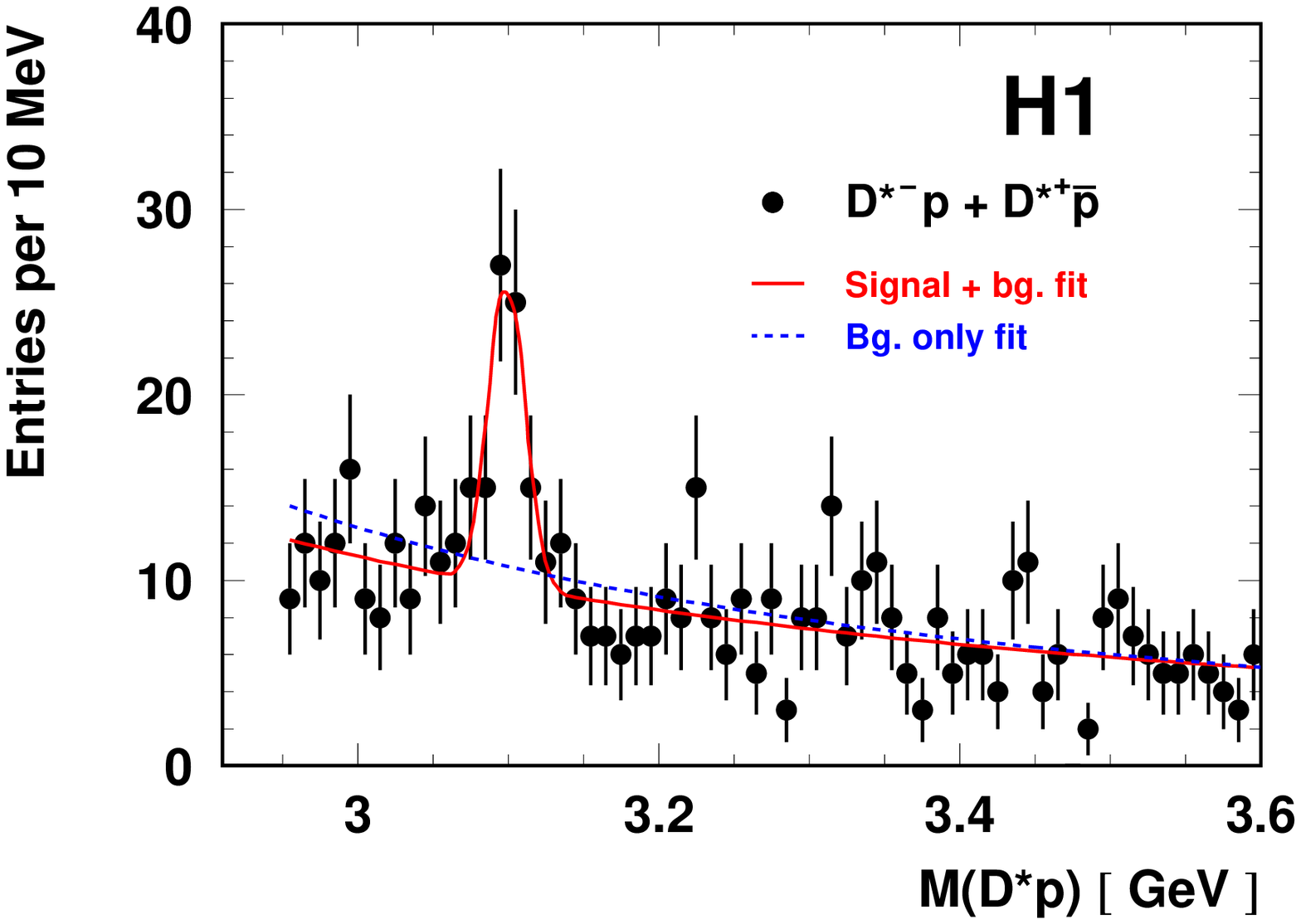}}
\put(230,0){\includegraphics[scale=0.38]{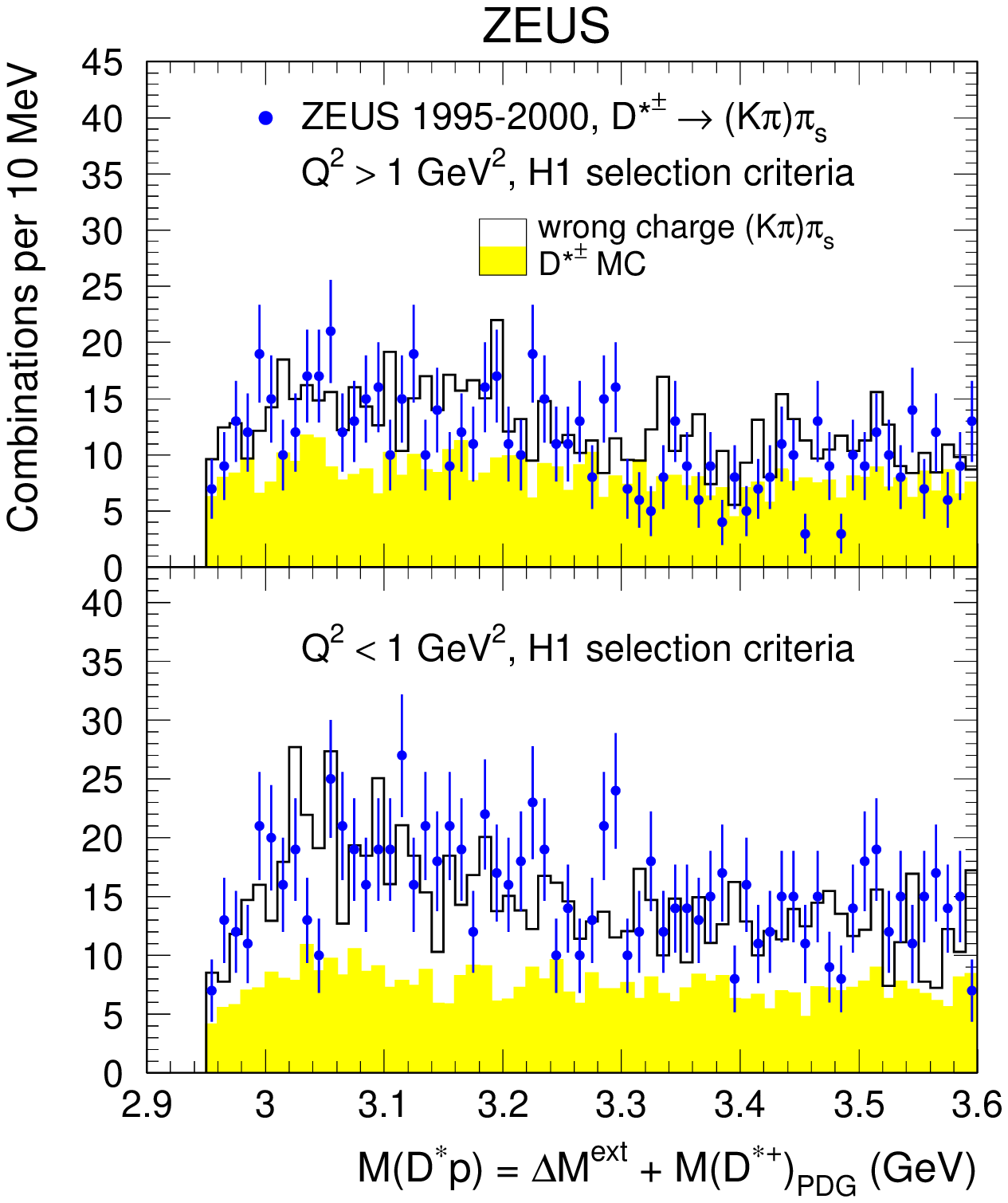}}
\put(188,144){\tiny(a)}
\put(356,151){\tiny(b)}
\put(356,79){\tiny(c)}
\end{picture}
\label{fig:DESY-04-164_6}
\caption{(a) $M(D^\ast p)=\Delta M+M(D^\ast)_{\rm PDG}$ distribution from opposite-charge 
$D^\ast p$ combinations in DIS obtained by H1, compared with the results of a fit in which both signal and background
components are included (solid line) and with the results of a fit in which only background 
is assumed (dashed line). (b) The distribution of $M(D^\ast p)$ for opposite-charge $D^\ast p$ combinations (points)
obtained by ZEUS using H1-like selection criteria in DIS with $Q^2>1$\,GeV$^2$ and (c) in photoproduction with
$Q^2<1$\,GeV$^2$. The histograms show a two-component model in which the wrong charge $(K\pi)\pi_s$ combinations
are used to describe the non-charm contribution and the inclusive $D^{*\pm}$ Monte Carlo simulation 
(shaded area) describes the contribution of real $D^{*\pm}$ mesons. The mass difference $\Delta M$ is defined as 
$M(K \pi \pi_s p)-M(K \pi \pi_s)$ and $M(D^\ast)_{\rm PDG}$ is the nominal $D^\ast$ mass.}
\end{figure}

The resonance is interpreted as 
an anti-charmed baryon with a minimal constituent quark composition of $uudd\bar{c}$, together with 
its charge conjugate.

\section{Search for a narrow anti-charmed baryonic state decaying to $D^{\ast\pm} p^\mp$ by ZEUS}

A similar search\cite{z3} has been performed by the ZEUS collaboration
in the $D^{\ast\pm} p^\mp$ invariant-mass spectrum using an integrated luminosity of $126\,$pb$^{-1}$.
The decay channels $D^{\ast+}\rightarrow D^0\pi^{+}_{s}\rightarrow(K^{-}\pi^{+})\pi^{+}_{s}$
and $D^{\ast+}\rightarrow D^0\pi^{+}_{s}\rightarrow(K^{-}\pi^{+}\pi^{+}\pi^{-})\pi^{+}_{s}$
(and the corresponding antiparticle decays) were used to identify $D^{\ast+}$ mesons in a combined DIS and 
photoproduction sample.
No resonance structure was observed in the mass spectrum in DIS as well as in photoproduction. The 
upper limit on the fraction of $D^*$ mesons originating from such a resonance decay is found to be 
0.23\% (95\% C.L.); the upper limit 
for DIS with $Q^2>1\,$GeV$^2$ is 0.35\% (95\% C.L.). To guarantee the highest possible comparability between 
the measurements of the H1 and the ZEUS collaborations the analysis has been repeated with requirements as close as 
possible to those of the H1 collaboration (Fig. 2 (b) and (c)); no signal was observed. The results
of the two collaborations are incompatible.

\section{Observation of $K_s^0K_s^0$ resonances in DIS}

Inclusive $K_s^0K_s^0$ production
has been studied\cite{z4} by the ZEUS collaboration in DIS using an integrated luminosity of $120$\,pb$^{-1}$. 
Two states are observed at masses of $1537^{+9}_{-8}$\,MeV and $1726\pm 7$\,MeV, 
as well as an enhancement around $1300$\,MeV (Fig. 3). The state at 
$1537$\,MeV is consistent with the well established $f^\prime$. The state at $1726$\,MeV may be 
the glueball candidate $f_0(1710)$. However, its width of $38^{+20}_{-14}$\,MeV
is narrower than the $125\pm10$\,MeV observed by previous experiments for the $f_0(1710)$.

\begin{figure}
\center{\includegraphics[scale=0.4]{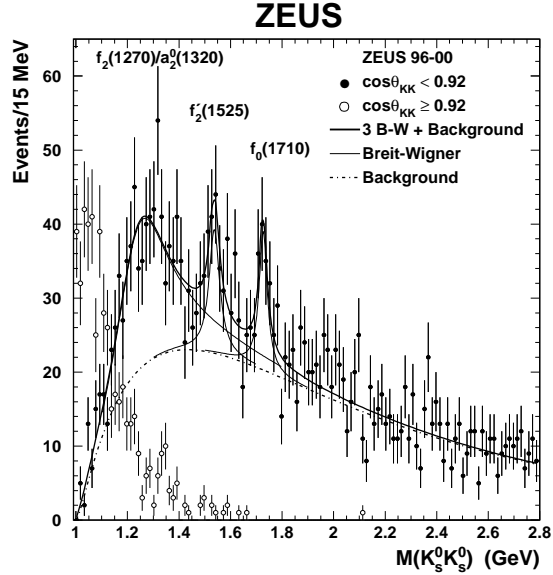}}
\label{fig:DESY-03-098_3}
\caption{The $K^0_sK^0_s$ invariant-mass spectrum for $K^0_s$ pair candidates with
$cos \theta_{K^0_sK^0_s} < 0.92$ (black points). The thick line represents a fit using
three  Breit-Wigners and a background parametrisation.} 
\end{figure}

\section{Conclusions}

Most of the recent spectroscopic measurements using the H1 and ZEUS detectors cover exotic 
resonances. The ZEUS collaboration has observed a resonance in $K^0_s p(\bar{p})$ which could be the $\Theta^+$ which has been
observed recently by several other experiments. The H1 collaboration has observed a narrow resonance in $D^\ast p$
which could be a candidate for a charmed pentaquark. However, this observation is not been confirmed by the 
ZEUS collaboration.


\section*{References}
\begin{thebibliography}{99}

\bibitem{pq1} LEPS Collab., T.~Nakano {\it et al.} \Journal{\PRL}{91}{012002}{2003};\\ 
                      SAPHIR Collab., J.~Barth {\it et al.} \Journal{\PLB}{572}{127}{2003};\\
		      CLAS Collab., V.~Kubarovsky {\it et al.} \Journal{\PRL}{91}{252001}{2003};\\
		      CLAS Collab., V.~Kubarovsky {\it et al.} \Journal{\PRL}{91}{032001}{2004}. \\
		      Erratum ibid, 049902
		      
\bibitem{pq2} DIANA Collab. V.V.~Barmir {\it et al.} \Journal{{\em Phys. Atom. Nucl.}}{66}{1715}{2003}; \\
              A.E.~Asratyan, A.G.Dolgolenko, M.A.Kubantsev, Preprint hep-ex/0309042 (2003); \\
              SVD Collab., A.~Aleev {\it et al.}, Preprint hep-ex/041024 (2004); \\
	      HERMES Collab., A.~Airapetian {\it et al.}, \Journal{\PLB}{585}{213}{2004}; \\
	      COSY-TOF Collab., M.~Abdel-Bary {\it et al.}, Preprint hep-ex/0403011 (2004);		      

\bibitem{z1} ZEUS Collab., S.~Chekanov {\it et al.}, \Journal{\PLB}{591}{7}{2004} 

\bibitem{na49} NA49 Collab., C.~Alt {\it eta al.}, \Journal{\PRL}{92}{042003}{2004}

\bibitem{z2} ZEUS Collab., S.~Chekanov {\it et al.}, \Journal{\PLB}{610}{212}{2005}

\bibitem{h1} H1 Collab., A.~Aktas {\it et al.}, \Journal{\PLB}{588}{17}{2004}

\bibitem{z3} ZEUS Collab., S.~Chekanov {\it et al.}, \Journal{{\em Eur. Phys. J.} C}{38}{29}{2004}

\bibitem{z4} ZEUS Collab., S.~Chekanov {\it et al.}, \Journal{\PLB}{578}{33}{2004}

\end{thebibliography}
\end{document}